\documentstyle[12pt,aaspp4,epsf]{article}

\tightenlines

\def\deg{$^{\circ}$}
\def\halpha{H$\alpha$}
\def\hbeta{H$\beta$}
\def\oiii{{\sc [O~iii]}}
\def\nii{{\sc [N~ii]}}
\def\sii{{\sc [S~ii]}}
\def\eg{e.g., }

\begin{document}

\centerline{(Accepted for Publication in the Sept. 1998 {\it PASP Research Notes)}}

\title{A Hidden Broad-Line Region in the Weak Seyfert 2 Galaxy NGC~788}

\author{Laura E. Kay}
\affil{Dept.\ of Physics and Astronomy, Barnard College, Columbia University,
NY, NY 10027
kay@calisto.phys.columbia.edu}

\and

\author{Edward C.\ Moran}
\affil{Dept.\ of Astronomy, University of California, Berkeley, CA 94720}


\begin{abstract}

We have detected a broad \halpha\ emission line in the polarized flux spectrum
of the Seyfert~2 galaxy NGC~788, indicating that it contains an obscured
Seyfert~1 nucleus.  While such features have been observed in $\sim 15$ other
Seyfert~2s, this example is unusual because it has a higher fraction of
galaxy starlight in its spectrum, a lower average measured polarization, and
a significantly lower radio luminosity than other hidden Seyfert~1s discovered
to date.  This demonstrates that polarized broad-line regions can be
detected in relatively weak classical Seyfert~2s, and illustrates why 
well-defined,
reasonably complete spectropolarimetric surveys at \halpha\ are necessary
in order to assess whether or not {\it all\/} Seyfert~2s are obscured
Seyfert~1s.

\end{abstract}


\section{INTRODUCTION}

One of the most pressing issues in the study of active galactic nuclei (AGNs)
is the nature of the nuclear activity in Seyfert~2 galaxies, which are defined
by the absence of the broad permitted optical emission lines that characterize
Seyfert~1 galaxies and quasars.  Initially, the spectroscopic differences
between type~1 and type~2 Seyferts were thought to arise simply from
differences in the location of the permitted line-emitting gas in their nuclei
(Weedman\markcite{16} 1977).  However, the obscuring torus model of 
Antonucci \& Miller\markcite{1} (1985),
which is based primarily on observations of the prototype Seyfert~2
galaxy NGC~1068, has provided a more sophisticated physical picture of Seyfert
galaxies. To account for their spectropolarimetry data, which revealed a
broad-line component in the polarized flux, Antonucci \& Miller suggested
that NGC~1068 contains a hidden Seyfert~1 nucleus whose continuum source and
broad-line clouds are completely blocked from view by a thick molecular torus.
Only a fraction of the radiation leaving the central source is
Thomson-scattered into our line of sight by a gas of warm electrons above
the torus.  By now, about 15 Seyfert~2 galaxies have had their hidden
broad-line regions (BLRs) revealed in polarized light 
(Miller \& Goodrich\markcite{10}
1990; Tran, Miller, \& Kay\markcite{12} 1992; Kay et al.\markcite{6} 1992; 
Tran\markcite{13,20} 1995a,b; Young\markcite{15} et al.
1996; Heisler, Lumsden, \& Bailey\markcite{4} 1997), 
indicating that they are really Seyfert~1 galaxies
for which we are located in an unfavorable viewing direction.  For these
objects, the orientation of the nucleus to our line of sight is principally
responsible for their spectroscopic properties.  The unification of the two
types of Seyfert galaxies in these cases is straightforward and elegant; for
this reason, it is frequently assumed to apply to {\it all\/} Seyferts.

The extrapolation of the obscuring torus model to all Seyferts has been
challenged on a number of occasions.  For example, recent studies have
indicated that the galaxian environments (De~Robertis, Yee, \& 
Hayhoe\markcite{2} 1998)
and dust properties (Malkan, Gorjian, \& Tam\markcite{19} 1998) 
of Seyfert~1 and Seyfert~2 galaxies
may differ; if so, these are circumstances which cannot be explained in
terms of geometrical effects on parsec scales.  Similarly, Moran
et al.\markcite{11} (1992)
reported that all known hidden Seyfert~1s have nuclear radio powers at
the high end of the Seyfert 20~cm radio luminosity function.  Evidence
for a hidden BLR should not correlate with radio power if the orientation
of the nucleus to our line of sight is the only parameter responsible for
the spectroscopic classification of Seyfert galaxies.  Using this reasoning,
Moran\markcite{11} et al. suggested that there may be two types of Seyfert~2 
galaxies,
hidden type~1 objects and ``true'' Seyfert~2s, which do not emit broad
emission lines.  But since the known hidden Seyfert~1s were not drawn from
the sample of Seyfert galaxies used to construct the radio luminosity
function, the connection between radio luminosity and the presence of
polarized broad emission lines needs to be re-investigated.

Even if the correlation between radio power and evidence for a hidden BLR
is verified, it may not provide a definitive test of the universality of the
unified Seyfert model.  Compared to the general population of Seyfert~2s,
hidden Seyfert~1s discovered by spectropolarimetry tend to have a lower
fraction of unpolarized host galaxy starlight in their optical spectra 
(Kay\markcite{7}
1994a), suggesting that they are more luminous objects relative to their
host galaxies.  If they are also more luminous Seyfert nuclei in absolute
terms, an apparent correlation between radio power and the existence of a
hidden BLR could arise from the fact that (1) the more luminous Seyferts
at optical wavelengths are also the brighter objects in the radio (e.g.,
Edelson\markcite{3} 1987; Whittle\markcite{17} 1992) and (2) spectropolarimetry 
is more effective at
uncovering polarized broad emission lines in objects with lower galaxy-light
fractions (Kay\markcite{7,8} 1994a,b).

In order to address these issues in an objective manner, we have initiated
a spectropolarimetric survey of a large 
distance-limited
sample of Seyfert~2
galaxies for which complete radio information is available.  Observations
of about half of the sample have been carried out to date.  In this
{\sl Research Note\/} we report the discovery of a hidden broad-line region
in the Seyfert~2 galaxy NGC~788, which is noteworthy because of the low
radio power and the starlight-dominated optical continuum this object
possesses.

\section{OBSERVATIONS} 

Observations of NGC~788 ($z=0.0136$) were obtained under clear skies on 1997 
September
29--30 using the 3m Shane reflector at Lick Observatory with the Lick
spectropolarimeter and the red beam of the KAST Spectrograph.
The 600~l~mm$^{-1}$ grating and $1200 \times 400$ Reticon CCD provided a
dispersion of 2.35~\AA/pixel and a resolution of 6--7~\AA\ (FWHM) over the
4600--7400~\AA\ range.  NGC~788 was observed for a total of two hours.
Each one-hour set consisted of four exposures with the halfwaveplate rotated
to 0\deg, 22.\negthinspace\deg5, 45\deg, and 67.\negthinspace\deg5.  Data
reduction and analysis were performed using the {\sl VISTA\/} software package,
and additional routines (Miller, Robinson, \& Goodrich\markcite{21} 1988).  All 
data
were summed in the Stokes parameters $Q$ and $U$, from which average
values for polarization $P$, postion angle $\theta$, and polarized flux
$P\times F$ were computed. 

\section{RESULTS}  

The total flux, polarization, 
polarization position angle, and the
corresponding polarized flux 
of NGC~788 are displayed as a function
of wavelength in Figure~1.  The measured polarization has not been corrected
for interstellar polarization or dilution by galaxy starlight. Even though
the average measured polarization $P$ is low---0.6\% at position angle
124\deg---it clearly increases across the \halpha\ line to a value of 1.2\%
while displaying little change in position angle.  Most interesting, though,
is the presence of a broad \halpha\ component in the {\it polarized\/} flux
spectrum.  The broad \halpha\ line peaks at the same wavelength as the narrow
component of \halpha\ in the direct-light spectrum and has an estimated
velocity width of 4800 km~s$^{-1}$ (FWHM).  Broad \hbeta\ is probably
present as well, but it is difficult to see in the low $S/N$ ratio polarized
flux spectrum.  The narrow lines of \oiii, \nii, and \sii\ are also visible
in polarized flux, suggesting that some of the measured polarization results
from transmission through a dust screen in either the Milky Way or the host
galaxy.   The nuclear radio source in NGC~788 is unresolved at $\sim$~$1''$
resolution (Ulvestad \& Wilson\markcite{14} 1989), 
so we cannot compare the polarization and radio source position angles.

It is instructive to compare the radio and optical continuum properties of
NGC~788 to those of the other hidden Seyfert~1s discovered to date.  To do
so, we draw upon the survey of Kay\markcite{7} (1994a), 
which contains a
magnitude-limited
sample of 50 Seyfert~2s, including NGC~788 and the ten hidden Seyfert~1s
studied by Tran\markcite{13} (1995a).  For most of the galaxies,
Kay\markcite{7} (1994a) estimated
the contribution of starlight to the optical spectrum.  Meaurements of their
core (arcsecond scale) 
nuclear radio luminosities have been culled from the literature by
Kay\markcite{9} et al. (1998).

The blue spectrum of NGC~788 (Kay\markcite{5} 1990) and that of the dwarf 
elliptical
galaxy M32 are displayed in Figure~2.  As evidenced by the strength of various
stellar absorption features, such as Ca~{\sc ii}~$\lambda$3968, the G~band
($\lambda$4302), Mg~{\sc i}~$b$~$\lambda$5176, and the Na~{\sc i}~D lines
($\lambda\lambda$5890,5896), it is clear that the Seyfert spectrum contains
a great deal of starlight and just a small amount of featureless continuum.
Near a wavelength of 4400~\AA , the fraction of the continuum arising from
galaxy starlight $F_{\rm g}$ is $\sim$~80\% (Kay\markcite{7} 1994a).  
This fraction is
rather typical for type~2 Seyferts in general ($\bar{F_{\rm g}}$ = 0.70),
but is much higher than the average value of 0.30 found for objects known to
have polarized broad emission lines (Kay\markcite{7} 1994a).\footnote
{Estimates of $F_{\rm g}$ reported for a given galaxy in different
studies vary because of a number of factors, such as the amount of galaxy
light falling on the spectrograph slit, the wavelength range considered in
the analysis, and how closely the type and reddening of the template galaxy
match those of the program object (Kay\markcite{7} 1994a). 
Although recent authors report
finding higher values of $F_{\rm g}$ for their small samples of Seyfert~2s,
(\eg Tran\markcite{13} 1995a; Cid~Fernandes\markcite{18} et al. 1998), 
we use measurements reported
by Kay\markcite{7} (1994a) since they were determined for a large number of 
objects
(both BLR and non-BLR Seyfert~2s) using a {\it consistent\/} procedure.}
In Figure~3 we have plotted the distribution of $F_{\rm g}$ for 47
Seyfert~2s in the Kay\markcite{7} (1994a) sample; the ten Seyfert~2s with 
hidden BLRs
studied by Kay\markcite{7} (1994a) and Tran\markcite{13} (1995a) 
(i.e., Mrk~3, Mrk~348, Mrk~463E,
Mrk~477, Mrk~1210, NGC~513, NGC~1068, NGC~7212, NGC~7674, and Was~49) are
shaded.  With the exception of NGC~513, NGC~788 has a higher starlight
fraction than the other hidden Seyfert~1s on the plot.  
NGC~513 actually has a prominent
broad \halpha\ component in its total-flux spectrum (Tran\markcite{14} 1995b), 
and is not,
strictly speaking, a type~2 Seyfert.

The radio luminosity of NGC~788 would appear to distinguish it from other
hidden Seyfert~1 nuclei as well.  With a core 20~cm radio power of $1.6 \times
10^{21}$ W~Hz$^{-1}$, NGC~788 falls at the faint end of the radio luminosity
function for nearby Seyferts (Ulvestad \& Wilson\markcite{14} 1989).  
As illustrated in Figure~4 (taken from Kay\markcite{9} et al. 1998), the
other Seyfert~2s known to have polarized broad emission lines are at least
30 times more luminous than NGC~788.

Several authors have suggested that the infrared colors of Seyfert~2s
may be related to the presence of a polarized hidden BLR in their nuclei
(Hutchings \& Neff\markcite{22} 1991; 
Heisler\markcite{4} et al.\ 1997).  For example, Heisler\markcite{4} et al.\
(1997) have claimed that Seyfert~2s with polarization evidence for a 
hidden BLR are more likely to have ``warm'' infrared colors (i.e.,
low $f_{60\mu{\rm m}} / f_{25\mu{\rm m}}$ ratios).  
Unfortunately, the {\it IRAS\/}
satellite obtained only an upper limit for the 25~$\mu$m flux density of
NGC~788, which does not provide a useful constraint on its infrared color.
Thus, we are unable to compare the infrared properties of NGC~788 to those
of other Seyfert~2s.

\section{CONCLUSIONS}

NGC~788 is a weak Seyfert~2 galaxy: its radio power is below average compared
to other classical Seyfert galaxies, and the amount of starlight in its
optical continuum is above average.  The detection of a polarized broad
\halpha\ emission line
in this galaxy demonstrates that hidden Seyfert~1 nuclei can be uncovered
in such unremarkable objects, which provides some interesting perspective
on the debate about the connection between core radio power and the presence
of of a hidden BLR.
Clearly, polarized broad emission lines are not exclusive to the most
powerful Seyfert~2s, contrary to the suggestion by Moran\markcite{11}
et al.\ (1992). On the other hand, it appears that
spectropolarimetry does not necessarily bias against the detection of
hidden BLRs in objects with starlight-dominated continua.  We cannot
draw more specific conclusions from just a single example; however, we are
confident that the results of our full survey, which will examine the degree
to which evidence for a hidden BLR correlates with isotropic properties of
Seyfert~2 galaxies, will provide important new insight into issue of Seyfert
unification.

\acknowledgments

We thank Andreea Petric for assistance with the observations. 
LK acknowledges support from NSF Faculty Early Development (CAREER)
grant AST-9501835.  ECM acknowledges support from NASA under grant
NAG5-3556.

\newpage

\clearpage

\begin{figure}
\epsscale{0.85}
\plotone{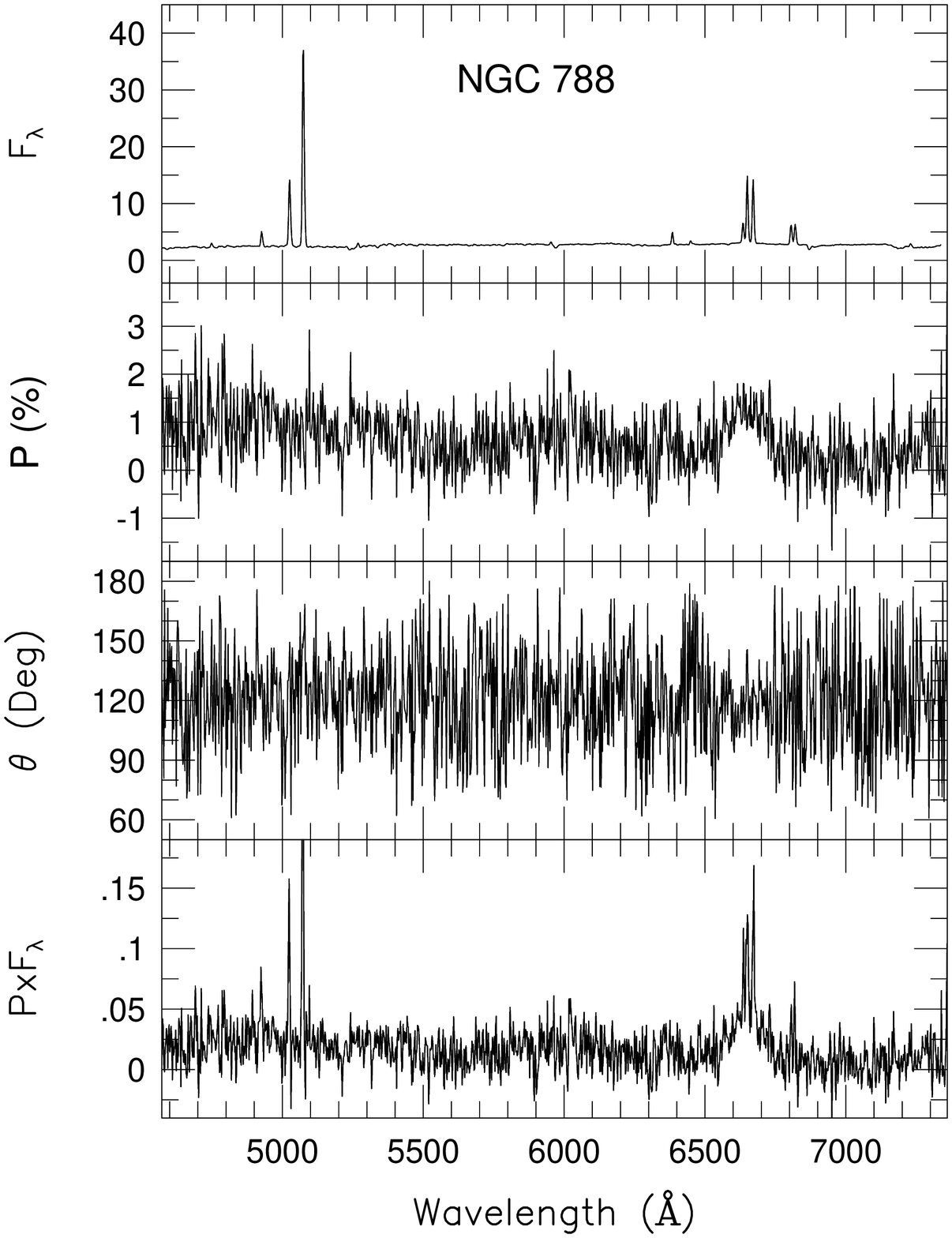}
\caption{Spectropolarimetry of NGC~788, uncorrected for reddening,
redshift, or galaxy starlight.  The flux (top panel) is in units of
$10^{-15}$ ergs s$^{-1}$ cm$^{-2}$ \AA$^{-1}$.  The second panel shows the
percent polarization (actually, the rotated Stokes parameter, which is
Stokes $Q$ rotated through the average polarization position angle).
The polarization position angle is displayed in the third panel, and the
bottom panel shows the corresponding polarized flux (the Stokes flux, which
is the product of the flux and the rotated Stokes parameter).}\label{fig1}
\end{figure}

\begin{figure} 
\plotone{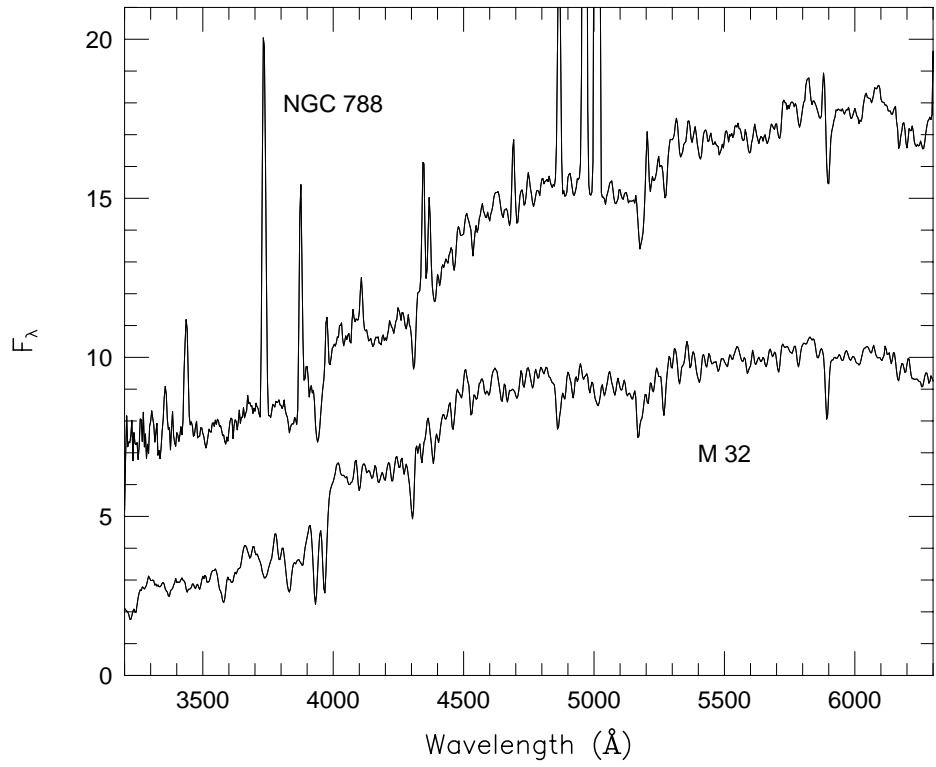} 
\caption{Blue spectra of NGC~788 and M~32, showing the high starlight
content in the NGC~788 spectrum. The data were obtained using the Lick 
3m with the UV Schmidt camera in 1988 and 1989 (Kay 1990).}\label{fig2} 
\end{figure} 
  
\begin{figure} 
\plotone{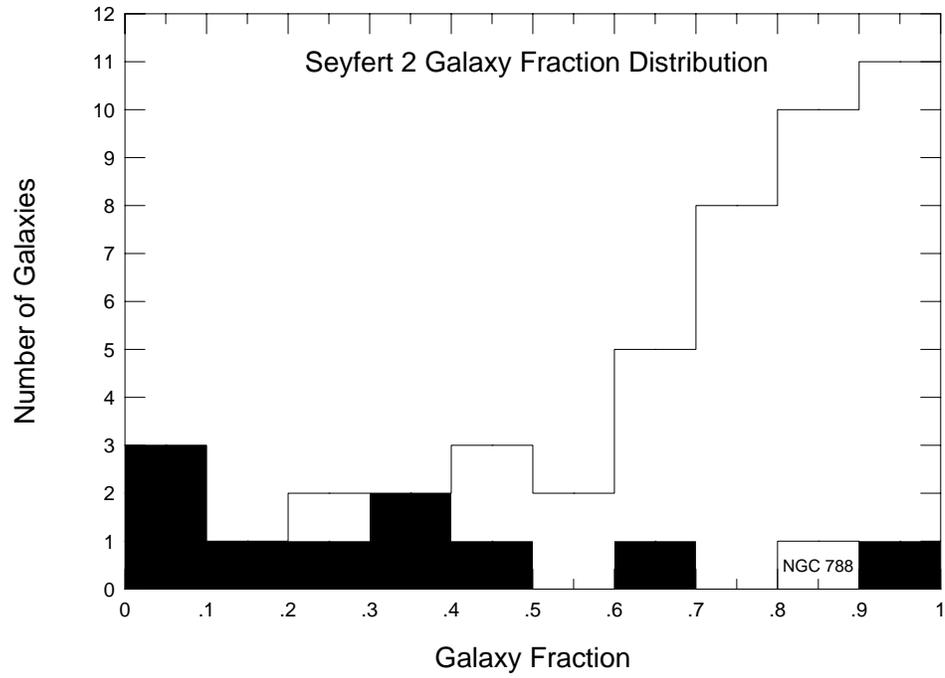} 
\caption{Distribution of $F_{\rm g}$ for the Seyfert~2s in Kay (1994a); 
shaded boxes represent the ten hidden BLR Seyfert~2s discussed in Tran 
(1995b).}\label{fig3}
\end{figure} 
 
\begin{figure} 
\plotone{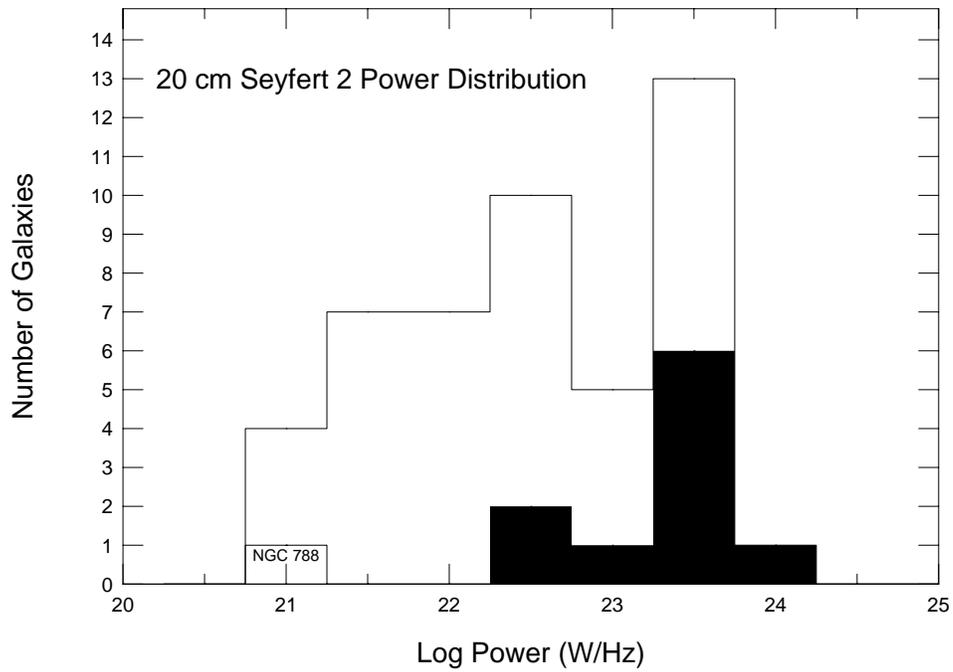}
\caption{Distribution of 20~cm radio power (in W~Hz$^{-1}$) for
the Seyfert~2s in Kay (1994a), as compiled by  Kay et al.\ (1998).
Shaded boxes are as in Fig.~3.  The luminosities
were calculated assuming $H_0 = 50$ km~s$^{-1}$~Mpc$^{-1}$ and
$q_0=0$.}\label{fig4}
\end{figure}

\end{document}